\documentstyle[12pt]{article}

\newcommand{\be}{\begin{eqnarray}}
\newcommand{\ee}{\end{eqnarray}}

\renewcommand{\vec}[1]{{\bf #1}}
\begin{document}

\vspace{2cm}

\begin{center}
{\Large Critical Ampitudes in Two-dimensional Theories.}
\vspace{1cm}

A.V. Smilga\\
\vspace{0.5cm}

{\it ITEP, B. Cheremushkinskaya 25, Moscow 117259, Russia}
\vspace{1cm}
\end{center}

\abstract{We derive exact analytical expressions for
the critical amplitudes $A_\psi$, $A_{gap}$ in the scaling laws for the fermion
condensate $<\bar \psi \psi>\ = \ A_\psi m^{1/3} g^{2/3}$
and for the mass of the lightest state $M_{gap} \ = \
A_{gap} m^{2/3} g^{1/3}$ in the Schwinger model with two
light flavors, $m \ll g$. $A_\psi$ and $A_{gap}$ are
expressed via certain universal amplitude ratios being calculated recently in 
TBA technique and the known coefficient $A_{\psi\psi}$ in the scaling law
$<\bar \psi \psi (x) \ \bar \psi \psi (0)>\ = \ A_{\psi\psi} (g/x)$ at the 
critical point.
Numerically, $A_\psi = -0.388\ldots,\ \ A_{gap} = 2.008\ldots$ . 
The same is done for the standard square lattice
Ising model at $T = T_c$. Using recent Fateev's results, we get 
$<\sigma_{lat}> = 1.058\ldots\  (H_{lat}/T_c)^{1/15}$ for the
magnetization and
$M_{gap} = a/\xi = 4.010\ldots (H_{lat}/T_c)^{8/15}$ for the
 inverse correlation length ($a$ is the lattice spacing). The
theoretical prediction for $<\sigma^{lat}>$ is in a perfect agreement
with numerical data. Two available numerical papers give the values of $M_{gap}$
 which differ from each other by a factor  $\approx \sqrt{2}$ .
 The theoretical result for  $M_{gap}$ agrees with one of them. }

\section{Introduction.}
A system with second order phase transition exhibits a
critical behavior at its vicinity. For example, the order
parameter $<\phi>$, which is zero at the phase transition
point $T = T_c$ when the corresponding symmetry is not
explicitly broken by external field, behaves as 
\be
\label{phih}
<\phi>_h({T =
T_c})\ = \ A_\phi \ h^{1/\delta}
\ee
 when external field is
present. The exponent $1/\delta$ is one of the critical
exponents. Another critical exponent $\mu$ appears in the scaling
behavior of the mass gap (the inverse correlation length)
 \be
  \label{Mh}
M_{gap}^h ({T=T_c}) \ =\  A_{gap}\ h^\mu
  \ee
The third critical exponent $\zeta$ we will be  interested in  
determines the power fall-off of the correlator $<\phi(\vec{x}) \phi(0)>$
 at $T = T_c$ at large distances
  \be
   \label{phicorr}
<\phi(\vec{x}) \phi(0)>_{h=0, T = T_c}\  \ = \ \
A_{\phi\phi} \ |\vec{x}|^{-(d-2 + \zeta)},\ \ \ \ \ \ |\vec{x}| \to \infty
  \ee
where $d$ is the spatial dimension.
 Three exponents $\delta$, $\mu$, and $\zeta$ are not independent
parameters but satisfy
two scaling relations (see e.g. \cite{Landau}) 
  \be
   \label{relind}
   \delta(d\mu - 1) \ =\ 1 \nonumber \\
\delta \mu(d-2+ \zeta) = 2
  \ee
Other critical exponents relate to the scaling behavior of the system at small
non-zero $|T - T_c|$ (see \cite{Landau} for the full list).

 The important fact is that many quite different physical systems can have
 the same values of critical exponents (they depend only on gross
symmetry features). This property is usually referred
to as {\it universality}. Universality is due to the fact that at small 
$|T-T_c|$
and small external fields $h$, the correlation length is high. At large
distances, a critical system ``forgets'' about pecilarities of microscopic 
interactions,
is scale invariant, and is described by an effective conformal field theory.

On the other hand, the values of the {\it critical amplitudes} like $A_\phi$
in (\ref{phih}) are not universal. Really, the order parameter $\phi$ and
the external field $h$ have distinct physical dimensions. The
coefficient $A_\phi$ also carries a dimension and depends on dimensionful 
constants in the microscopic hamiltonian.

It is well known, however, that certain dimensionless combinations of 
critical amplitudes exist which, like exponents, are determined by only a 
large--distance behavior of the system and are universal \cite{ratios}. 
Actually, any scaling relation between exponents is associated with a certain
universal ratio.
 
As an illustration,  consider the shift in
free energy density of the critical system at $T = T_c$ due to the
presence of external field $h$. We have
  \be
  \label{Fh}
  \Delta F(h) \ =\ bT_c M_{gap}^d
  \ee
  where $b$ is a dimensionless numerical coefficient which is
universal. Bearing in mind that $<\phi>_h = \partial F(h)/\partial h$,
we derive the first relation in (\ref{relind}) and, simultaneously,
that the dimensionless ratio
  \be
  \label{r1}
  r_1 \ = \ \frac{A_\phi}{T_c A_{gap}^d} = b\mu d
  \ee
  is universal.

The second universal relation in (\ref{relind}) is derived considering the 
correlator of
order parameters $<\phi(\vec{x}) \phi(0)> $ at $T=T_c$ in the presence
of external field. When $h$ is small, the correlator exhibits first a 
power fall--off as in (\ref{phicorr}). Then at 
$|\vec{x}| \sim \xi = M_{gap}^{-1}$ the
behavior of the correlator is modified. Asymptotically, it tends to
$<\phi>_h^2$. Preasymptotic terms decay exponentially $\sim
\exp\{-M_{gap}|\vec{x}|\}$. When the correlation length is high, the behavior
of correlator at the distances $|\vec{x}| \sim \xi$ should not depend
on the details of microscopic interactions but only on the dynamics of
the effective conformal theory describing a critical system in the scaling
regime. 
In other words, 
$$<\phi(|\vec{x}| = \xi) \ \phi(0)> \ = \ c<\phi>^2$$
where $c$ is a universal constant.
We derive thereby a second relation in (\ref{relind}) and also that
the dimensionless ratio
  \be
  \label{r2}
  r_2 \ = \ \frac {A_\phi^2}{A_{\phi \phi} A_{gap}^{d-2+\zeta} } \ = c 
 \ee
  is universal.

 In physical 3--dimensional systems, the values of critical
exponents  and universal critical ratios are calculated numerically as a 
series over the
parameter $\epsilon = 4 - d$ \cite{Wilson,ratios}. In many two--
dimensional statistical systems, {\it both} can be
determined analytically (the mathematical reason for that is that
conformal group in two dimensions is much richer and imposes much
stringer constraints on the behavior of the system that at $d \geq 3$). 

The values of exponents for the Ising model and many other exactly solved 
two--dimensional critical systems were known for a long time. Recently, it 
has become clear that many 2D critical systems in some vicinity of critical 
point (in particular, the Ising model at $T = T_c$ in weak external 
magnetic field)
are described by exactly integrable two--dimensional field theories with a known
$S$--matrix. The ingenious Thermodynamic Bethe Ansatz (TBA) technique has been
developed \cite{AlZam,Klassen} which allowed one to evaluate the universal 
ratios $r_{1,2}$ analytically.

With the ratios $r_{1,2}$ at hand, one only need to know one of the amplitudes
$A_{\phi}$, $A_{gap}$, $A_{\phi \phi}$  to determine two others.
In this paper, we exploit this fact and discuss two examples of 
two-dimensional exactly
solved models with second order phase transition --- the Ising model on the 
square lattice
and the Schwinger model with two fermion flavors. In both cases, the critical 
amplitude $A_{\phi \phi}$ (which  {\it is} not universal and depends on a 
particular form of the microscopic hamiltonian) has been determined 
from independent premises.

Thereby the amplitudes $A_{\phi}$ and $A_{gap}$ can also be determined.

 \section{Ising Model.}
 \setcounter{equation}0
 
A classical example of an exactly solved two--dimensional critical
system is the Ising
model. The hamiltonian of the model reads
 \be
 \label{HIsing}
{\cal H} \ =\ -J \sum_{ij} ( \sigma_{ij} \sigma_{i+1,j}\ +\
\sigma_{ij} \sigma_{i,j+1})\ -\
 H \sum_{ij} \sigma_{ij}
  \ee
  The corresponding partition function is
  \be
  \label{ZIsing}
  Z\ = \ {\rm Tr} \exp\{-{\cal H}/T\} \ = \nonumber \\
  \sum_{\{\sigma_{ij}\}}
  \exp \left \{\beta  \sum_{ij} ( \sigma_{ij} \sigma_{i+1,j}\ +\
\sigma_{ij} \sigma_{i,j+1}) \ +\ h \sum_{ij} \sigma_{ij} \right \}
  \ee
  where $\beta = J/T, \ h = H/T$.
It was known for a long time that at $\beta\ =\ \beta_c \ = \ 
\frac 12 \ln(1+\sqrt{2})$
  \be
 \label{Isscale}
<\sigma>_h \ =\ A_\sigma \ {\rm sign}(h) \ h^{1/15}
\nonumber \\
M_{gap} \ =\ A_{gap} \ h^{8/15} \nonumber \\
<\sigma(\vec{x}) \sigma(0)>_{h=0}\ = \ \frac{ A_{\sigma \sigma}}
{|\vec{x}|^{1/4}}
\ee
The coefficient $A_{\sigma\sigma}$ was determined some time ago  by
direct evaluation of the lattice correlator in the theory (\ref{ZIsing})
  \be
  \label{McCoy}
  <\sigma_{NN} \ \sigma_{00}> \ \sim \frac C {N^{1/4}},\ \ \ \ \ \ \ \ N
\to \infty
  \ee
  where $C = .645\ldots$\ is a known trancedental constant \cite{McCoy}.
  From this, one easily gets
  \be
  \label{Ass}
  A_{\sigma\sigma} = C 2^{1/8} = .703\ldots
  \ee
  where distance is measured in the units of lattice spacing which we
set to one in the subsequent discussion. The constant
$A_{\sigma\sigma}$ depends on the particular form of the hamiltonian
(\ref{HIsing}) and is not universal - its value is different, say, on a 
triangle lattice or in a model with not only nearest neighbors interaction. 

As was already noted, the  analytical 
determination of the universal ratios (\ref{r1}, \ref{r2})   has become 
possible only recently after a
beautiful A.Zamolodchikov's work who described the Ising model
at critical temperature and at weak external magnetic field  (so that the 
correlation length is much larger than the lattice spacing) 
as a perturbed conformal field theory 
 \be
  \label{Sh}
  S_{\rm Ising}(h) \ = \   S_{\rm Ising}(0)\ - \ h^{CFT} 
  \int \sigma^{CFT}(\vec{x})
\   d^2x
 \ee
 where $\sigma^{CFT}$ is the conformal spin field normalized such that
 \be
 \label{norm}
 <\sigma^{CFT}(\vec{x}) \ \sigma^{CFT}(0)>_{h^{CFT} = 0}\ =\ 
\frac 1{|\vec{x}|^{1/4}}
 \ee
so that $A^{CFT}_{\sigma \sigma} = 1$. 
Zamolodchikov found out that the model is exactly integrable 
involving an infinite
number of conserved charges. The spectrum of the model includes 8 states with 
definite peculiar mass ratios. These states scatter on each other without 
reflection and the $S$--matrix is  exactly calculable  \cite{ZamSash}. 

TBA technique \cite{AlZam} allows one to find the universal ratios $r_{1,2}$.
The ratio $r_1$ in the wide class of theories has been found in \cite{Klassen}.
Speaking precisely, the coefficient $b$ entering the free energy density
(\ref{Fh}) was evaluated. The general formula is
  \be
  \label{bphi}
b =  - \frac 1{2\phi_{11}^{(1)}}
  \ee
where $\phi_{11}^{(1)}$ is a certain constant extracted from the high energy 
asymptotics of the scattering amplitude of the state with the lowest mass:
$$ \phi_{11}^{(1)} \ =\ i \lim_{\theta \to \infty} \ e^\theta \ \frac d 
{d\theta} S_{11}(\theta) $$
($\theta$ is the rapidity). For the Ising model,
 \be
  \label{bIsing}
b \ =\ - \frac 1{16\sqrt{3} \cos (\pi/30) \sin (\pi/5)}
  \ee
The ratio $r_1$ is then given by  Eq. (\ref{r1}). The ratio $r_2$  (actually, 
the combination  $r_1^2/r_2$) has been determined recently by Fateev  
\cite{Fateev}. The results can be presented as explicit expressions for  the 
amplitudes
 $A_\sigma$ and $A_{gap}$ when the normalization convention (\ref{norm}) is 
chosen (see also \cite{DelGuid}). The result is
  \be
  \label{Fat}
<\sigma^{CFT}> \ =\  \frac 8{15} \left( \frac {4\pi^2
\Gamma^2(\frac {13}{16}) \Gamma(\frac 34)}{\Gamma^2(\frac 3{16})
 \Gamma(\frac 14)}
\right)^{8/15} \frac {\sin (\frac \pi 5)}{\sin (\frac \pi 3) \sin
(\frac {8\pi}{15})} \left( \frac {\Gamma(\frac 15)}{\Gamma(\frac 23)
\Gamma(\frac 8{15})} \right)^2 \left(h^{CFT}\right)^{1/15}  \nonumber \\ 
= \  1.277\ldots \left(h^{CFT}\right)^{1/15}
\nonumber \\
M_{gap} \ = \  \left( \frac {4\pi^2 \ \Gamma^2(\frac {13}{16})\ 
\Gamma(\frac 34)}{\Gamma^2(\frac 3{16})\ \Gamma(\frac 14)} \right)^{4/15}\
 \frac {4 \sin (\frac \pi 5)\  \Gamma(\frac 15)}{\Gamma(\frac 23)\ 
\Gamma(\frac 8{15})} \left(h^{CFT}\right)^{8/15} \nonumber \\
 = \ 4.404\ldots \left(h^{CFT}\right)^{8/15}
\ee

To find out the critical amplitudes relating the physical spin
expectation value and the physical correlation length to the physical
magnetic field in the standard Ising model on square lattice, one should 
take into account the difference in
 normalizations of $\sigma^{lat}$ and $h^{lat}$ vs $\sigma^{CFT}$
  and $h^{CFT}$. Comparing (\ref{ZIsing}) and (\ref{McCoy}) with 
(\ref{Sh}) and (\ref{norm}), we obtain
 \be
 \label{latCFT}
 \sigma^{lat} \ = \ A_{\sigma \sigma}^{1/2}  \sigma^{CFT}, \nonumber \\
 h^{lat} \ = \  A_{\sigma \sigma}^{-1/2} h^{CFT}
 \ee
 From this and Eq. (\ref{Fat}) a final result can be derived
 \be
 \label{siglat}
 <\sigma^{lat}> \ = \ 1.277\ldots (A_{\sigma \sigma})^{8/15}
 \left(h^{lat}\right)^{1/15}\ =\ 1.058\ldots \
\left(h^{lat}\right)^{1/15}
  \ee
  \be
  \label{colat}
M_{gap}\ =\  1/\xi \ = \ 4.404\ldots (A_{\sigma \sigma})^{4/15} 
\left(h^{lat}\right)^{8/15}
\ =\ 4.010\ldots \left(h^{lat}\right)^{8/15}
 \ee
 The theoretical result for $<\sigma^{lat}>$ perfectly agrees with 
the available numerical data  $<\sigma^{lat}> \ =\  0.999(1) 
(h^{lat}/\beta_c)^{1/15}$  \cite{Lauwers} and  $<\sigma^{lat}> \ =\  1.003(2) 
(h^{lat}/\beta_c)^{1/15}$ \cite{Destri}. The prediction
(\ref{colat}) for the mass gap agrees well with the numerical result 
$\xi = 0.38(1)  (h^{lat}/\beta_c)^{-8/15}$ \cite{Lauwers} but dramatically   
{\it disagrees} (by the factor $ \sim \sqrt{2}$ ) with the numerical 
result $M_{gap} = 1.839(7) (h^{lat}/\beta_c)^{8/15}$ as given in \cite{Destri}.
Thereby these two numerical works contradict to each other at this point.  
 One can further notice that the theoretical value of the universal constant 
$b$ as quoted in \cite{Destri} is twice as large as it should be [ the 
factor 8 instead of 16 in Eq. (\ref{bIsing}) ].
The reasons of this disagreement are  not clear.

\section{Schwinger model with $N_f = 2$.}
\setcounter{equation}0

Our remark is that the critical coefficients can also be
determined along similar lines in another exactly solved
model with critical behavior --- the Schwinger model with
two fermion flavors. The Euclidean lagrangian of the model is
  \be
  \label{SM}
{\cal L}\ = \frac 12 F^2 \ -i\ \sum_{f=1,2} \bar \psi_f
\gamma^\mu (\partial_\mu - igA_\mu) \psi_f \ +\ m
\sum_{f=1,2} \bar \psi_f \psi_f
  \ee
where $F = F_{01}$ and $\gamma_\mu$ are anti-hermitian. All fields live 
in 1+1 dimensions. The
coupling constant $g$ has the dimension of mass. At $m = 0$,
the theory enjoys the chiral $SU_L(2) \otimes SU_R(2)$ symmetry
much like as standard $QCD_4$. The corresponding order
parameter is the fermion condensate $<\bar \psi_1 \psi_1>\  =\ 
<\bar \psi_2 \psi_2>$. The appearance of non-zero condensate
would break spontaneously chiral symmetry. Spontaneous
breaking of a continuous symmetry is not possible, however,
in 1+1 dimensions \cite{Coltheor}. Hence the condensate is
zero when $m=0$.

In spite of the absence of the ordered phase, it has been
shown that the second order phase transition still occurs in
the massless multiflavor Schwinger model at {\it zero
critical temperature} \cite{Ver}. That means that at $T=0$
and at small positive temperatures the system behaves much
like a critical system at the phase transition point or
slightly above. Also  at zero temperature and at small non-zero fermion
mass, the correlators, the fermion
condensate, and mass gap exhibit a critical behavior
  \be
  \label{critSM}
<\bar \psi_1 \psi_1>_m \ =\ -A_\psi m^{\frac {N_f-1}{N_f+1}}
g^{\frac 2{N_f+1}} \nonumber \\
M_{gap}(m)\ = \  A_{gap}  m^{\frac {N_f}{N_f+1}} g^{\frac
1{N_f+1}} \nonumber \\
<\bar \psi_1 \psi_1 (x) \ \bar \psi_1 \psi_1(0)>_{m=0}
\ =\  A_{\psi \psi} \frac {g^{2/N_f}}{x^{2-2/N_f}},\ \ \ \
gx \gg 1 
  \ee
where $N_f$ is the number of light flavors.
\footnote{To avoid confusion, note that the universal ratio $r_1$ is defined 
now without the factor $(T_c)^{-1}$  as in Eq.(\ref{r1}) here. 
The matter is, in statistical systems everything is usually defined via free energy
density $F \ =\ -T/V \ln Z$ while in a field theory a more natural quantity is
the vacuum energy density $\epsilon_{vac} = - 1/V^{Eucl} \ln Z$. Non-zero 
temperatures in Schwinger model would correspond to the same theory defined on an 
Euclidean
2-dimensional cylinder whereas changing the "statistical temperature" means changing
coupling constants or adding extra terms in the Euclidean theory framework. }
 A small fermion
mass $m \ll g$ plays here the same role as the small magnetic
field $h$ in the Ising model. It breaks explicitly the
chiral symmetry
 $SU_L(N_f) \otimes SU_R(N_f)$ of the massless theory down
to $SU_V(N_f)$. The values of critical exponents in
(\ref{critSM}) have been calculated analytically for any
$N_f$ in \cite{Coleman}--\cite{cond}. The critical coefficient
$A_{\psi \psi}$ for the fermion correlator can also be
easily determined \cite{Abada} using the fact that the corresponding path
integral has a Gaussian form and can be calculated exactly.
We have
  \be
  \label{A2psiN}
 A_{\psi \psi} \ =\ \frac {e^{2\gamma/N_f}}{2\pi^2} \left( \frac {N_f}{4\pi}
 \right)^{1/N_f}
  \ee
  where $\gamma = 0.577\ldots$ is the Euler constant.
The coefficient $A_{\psi \psi}$ is sensitive to the short distance
region of the theory $x \sim 1/g$ (the full coefficient
 $A_{\psi \psi}\ g^{2/N_f}$
 depends explicitly on the intrinsic mass scale
 $g$) and is not universal.

In this paper we determine the coefficients $A_\psi$ and
$A_{gap}$ in the case $N_f = 2$.
Our starting point is the abelian bosonization procedure of
\cite{Coleman}. We identify
  \be
  \label{boson}
i \bar \psi_1 \gamma_\mu \psi_1 \ \equiv \ \frac 1{\sqrt{\pi}}
\epsilon_{\mu\nu} \partial_\nu \phi_1,\ \ \ \ \ \ 
m \bar \psi_1 \psi_1 \ \equiv \ -C \cos \sqrt{4\pi} \phi_1
\nonumber \\
i \bar \psi_2 \gamma_\mu \psi_2 \ \equiv \ \frac 1{\sqrt{\pi}}
\epsilon_{\mu\nu} \partial_\nu \phi_2,\ \ \ \ \ \ 
m \bar \psi_2 \psi_2 \ \equiv \ -C \cos \sqrt{4\pi} \phi_2
  \ee
The original theory (\ref{SM}) is equivalent to the bosonic
theory
  \be
  \label{SMbos}
{\cal L} \ = \  \frac 12 F^2 + \frac 12 (\partial_\mu
\phi_1)^2 +
\frac 12 (\partial_\mu \phi_2)^2
+ ig \frac F {\sqrt{\pi}} (\phi_1 + \phi_2)  \nonumber \\
- C \left( \cos \sqrt{4\pi} \phi_1 \  +\  \cos \sqrt{4\pi} \phi_2 \right)
  \ee
in a sense that it has the same spectrum and that all
correlators of fermion currents in the theory (\ref{SMbos})
coincide with the correlators of the corresponding bosonic
currents in the theory (\ref{SMbos}) \cite{boson}.
We can integrate now over $F$ and arrive at the following bosonic
lagrangian involving only physical degrees of freedom
 \be
 \label{SMphi}
 {\cal L} \ = \ \frac 12 (\partial_\mu \phi_+)^2 \ +\ 
  \frac 12 (\partial_\mu \phi_-)^2 \ + \ \frac {g^2}\pi \phi_+^2 \ 
- \ 2C \cos (\sqrt{2\pi} \phi_+) \cos (\sqrt{2\pi} \phi_-)
 \ee
 where $\phi_\pm = (\phi_1 \pm \phi_2)/\sqrt{2}$. 
 
 If the original fermion theory is massless, $C = 0$ and we have 
the theory of two free bosonic fields. One of them ($\phi_+$) has the mass
 \be
 \label{muSM}
 \mu_+^2 \ =\ \frac {2g^2}\pi
 \ee
 and the other ($\phi_-$) is massless. The absence of the mass gap
means that the correlation length of the system is infinite. That
results in the power fall-off of the correlator of order parameters 
$C_{11}(\vec{x}) \ =\ <\bar \psi_1 \psi_1 (\vec{x}) \ \bar \psi_1 \psi_1(0)> \ 
\sim \ g/|\vec{x}|$ at
large distances which  characteristizes a critical system at the phase 
transition point.
If $m$ is non-zero but small,
$C$ is also non-zero and small and the fields begin to interact. We are
interested in the dynamics of the system at large distances and small energies.
Then the heavy field $\phi_+$ decouples (it freezes down at the value
$\phi_+ = 0$) and the system is described by the effective lagrangian involving
only the light field $\phi_-$:
 \be
 \label{Lphim}
 {\cal L}^{eff}\ =\   \frac 12 (\partial_\mu \phi_-)^2 - 2C
 \cos (\sqrt{2\pi} \phi_-)
  \ee
 In this limit,
 \be
 \label{psi12}
m\bar\psi_1 \psi_1 \ \equiv \ m\bar \psi_2 \psi_2 \ \equiv \ -C 
\cos (\sqrt{2\pi} \phi_-)
 \ee
and the correlators $C_{11}(\vec{x}) $
and $C_{12}(\vec{x}) \ =\ <\bar \psi_1 \psi_1(\vec{x}) \ 
\bar \psi_2 \psi_2(0)>$ coincide.
\footnote{Note that, in the fermion language, the correlator $C_{11}(\vec{x})$
 is saturated 
by topologically trivial gauge fields while the correlator $C_{12}(\vec{x})$
 --- by the fields belonging to 1-instanton topological sector.}
 
 The lagrangian (\ref{Lphim}) describes the sine-Gordon model which, like (\ref{Sh}), 
can be treated as
a perturbation of the conformal theory ${\cal L} = (\partial_\mu \phi_-)^2/2$.
Like (\ref{Sh}), the model (\ref{Lphim}) is exactly solved and can be
analyzed along similar lines. Actually, this analysis is much simpler
here. The sine-Gordon model is the first known example of a non-trivial
non-linear theory where exact $S$--matrix has been constructed \cite{Zamsin}.
 The spectrum and the free energy density of the sine-Gordon model have
been recently found by Al. Zamolodchikov \cite{ZamLesh}. He studied the
model
  \be
   \label{SG}
   {\cal L}^{SG} \ =\ \frac 12 (\partial_\mu \phi)^2 \ - \ 2C \cos (\beta
\phi)
\ee
at arbitrary coupling $\beta$ assuming the normalization
 \be
 \label{normSG}
 <\cos [\beta \phi(\vec{x})]\  \cos [\beta \phi(0)]>_{C=0} \ = \ \frac
1{2|\vec{x}|^{\beta^2/2\pi}} 
 \ee
The spectrum of the model involves a soliton, an antisoliton and some
number of the soliton-antisoliton bound states. For $\beta =
\sqrt{2\pi}$ (the case we are interested in) there are just two such
bound states. One of them is has the same mass $M_{gap}$ as the solitons
 so that  these three states form an isotopic triplet (recall that the
original fermion model (\ref{SM}) had the isotopic $SU(2)$ symmetry and 
so should
its bosonized version), and the other one has the mass 
 $M_{gap} \sqrt{3}$ and is an isotopic singlet.
Actually, Sine--Gordon model   with $\beta = \sqrt{2\pi}$  is simpler than a 
theory with an arbitrary $\beta$. It enjoys a pure elastic scattering matrix.
 Its associated Lie algebra is $D_4^{(1)}$ (cf. e.g. Table 1 in \cite{Klassen}).
As was  pointed out in \cite{Affleck} and recently in \cite{Hoso,Hoso1}, 
it belongs to the same 
universality class as the antiferromagnetic quantum spin chain \cite{Affleck1}.

 The Zamolodchikov's result for the
mass gap $M_{gap}$ in the model  (\ref{SG}) with  $\beta =
\sqrt{2\pi}$ reads
  \be
   \label{gapSG}
   M_{gap} \ = \  \frac{2\pi^{1/6}\  \Gamma^{2/3}(\frac 34)
\Gamma(\frac 16)}{\Gamma^{2/3}(\frac 14)\ \Gamma(\frac 23)}\ C^{2/3}
   \ee
   In order to express $M_{gap}$ via physical parameters $m$ and $g$, we
have to fix the coefficient $C$.  Using the result
(\ref{A2psiN}) with $N_f =2$ for the physical fermion correlator, 
the property (\ref{psi12}) 
and the definitions (\ref{boson}), (\ref{normSG}), we obtain
  \be
  \label{Cgm}
  C \ =\ \frac{mg^{1/2} e^{\gamma/2}}{2^{1/4}\pi^{5/4}}
\ee
Substituting it in (\ref{gapSG}), we finally derive
 \be
 \label{gapSM}
 M_{gap} = m^{2/3} g^{1/3} 2^{5/6} e^{\gamma/3}
 \left[ \frac {\Gamma(\frac 34)}{\pi \Gamma(\frac 14)} \right]^{2/3} \ \frac
 {\Gamma(\frac 16)}{\Gamma(\frac 23)} \ =\ 2.008\ldots m^{2/3} g^{1/3}
   \ee
To find the coefficient $A_\psi$, we use the expression for the vacuum energy
density of the model   derived in \cite{Klassen,Vega}. For  $\beta =
\sqrt{2\pi}$ it reads
 \be
 \label{Evac}
 \epsilon_{vac} \ =\ -\frac {M_{gap}^2}{4\sqrt{3}}
 \ee
 Differentiating it over fermion mass $m$, we obtain the expression for
the fermion condensate
  \be
   \label{condSM}
   <\bar \psi_1 \psi_1>_{vac} \ = \    <\bar \psi_2 \psi_2>_{vac} \ = \ 
   \frac 12 \frac {\partial \epsilon_{vac}}{\partial m} \nonumber \\
   = \ - m^{1/3} g^{2/3} \frac {2^{2/3} e^{2\gamma/3}}{3\sqrt{3} \pi^{4/3}}
   \left[ \frac {\Gamma(3/4)}{\Gamma(1/4)} \right]^{4/3}
   \left[ \frac {\Gamma(1/6)}{\Gamma(2/3)} \right]^2\ =\ -0.388\ldots
    m^{1/3} g^{2/3}
   \ee

The mass gap in the Sine--Gordon model was evaluated earlier by quasiclassical 
methods.
In \cite{Dashen} the lagrangian of the model was chosen in the form
  \be
  \label{SGold} 
{\cal L} \ = \ \frac 12 (\partial_\mu \phi)^2 - \frac {\mu^2}{\beta^2} 
{\cal N}_\mu\cos (\beta \phi)
  \ee
where $\mu$ is the meson mass in the weak coupling (small $\beta$) limit and
 ${\cal N}_\mu$ is the normalization ordering prescription with respect 
to that mass. 
Quasiclassically,
the soliton mass is \cite{Dashen}
 \be
 \label{msol}
M_{sol} \ =\ \left( \frac 8{\beta^2} \ -\ \frac 1\pi \right) \mu
 \ee
For $\beta = \sqrt{2\pi}$, $M_{sol} = 3\mu/\pi$.
To compare it with the exact Zamolodchikov's result (\ref{gapSG}) and its 
corollary 
(\ref{gapSM}), we have to relate $\mu$ and $C$. It can be easily done comparing 
Eq.(\ref{normSG}) with
 \be
  \label{norm1}
<{\cal N}_\mu \cos [\sqrt{2\pi} \phi(x)]\ \ {\cal N}_\mu \cos [\sqrt{2\pi} \phi(0)]> \ =
\ \cosh[2\pi D(\mu, x) ] \nonumber \\
 \sim \cosh[-\gamma - \ln(\mu x/2) ]  \sim \ \frac {e^{-\gamma}}{\mu x}
  \ee
for small $\mu x$. We have 
$$ \mu \ = \ (2\sqrt{2} \pi e^{\gamma/2} C)^{2/3} $$
Substituting it in Eq.(\ref{msol}) and taking into account (\ref{Cgm}),
 we would get
  \be
A^{WKB}_{gap} \ =\ 2.07\ldots
  \ee
We see that the WKB result turned out to be rather close to our exact result 
(\ref{gapSM}). The difference in just 3\%. May be, there is no wonder that the 
accuracy of quasiclassical analysis is so high. Notice that the quasiclassical 
ratio $M_{sol}/\mu \ =\ 3/\pi$ is also very close to 1 --- 
the theoretical prediction for the
ratio of the soliton mass to the lowest breather mass for $\beta = \sqrt{2\pi}$.  
Note also that if we would estimate the mass gap as the classical mass $\mu$ 
of the basic 
Sine--Gordon boson rather than as $M_{sol}$ ( that was in fact done in \cite{Hoso}),
 we would get $A_{gap}^{WKB} =2.16\ldots$  and 
the agreement would be somewhat worse. 
%  The quantum corrections were calculated 
%numerically in \cite{JacSG}. For $\beta = \sqrt{2\pi}$ they were found to increase 
%the mass by about 15\%.
 In    \cite{Hoso} also the critical amplitude for the fermion condensate was 
estimated by quasiclassical methods. The result
  \be
  \label{Hosam}
 <\bar\psi \psi> \ = \ - \left( \frac {e^{4\gamma}}{2\pi^4} \right)^{1/3} 
m^{1/3} g^{2/3} \ = \ - .37\ldots \ m^{1/3} g^{2/3}
  \ee
agrees rather well with the exact formula (\ref{condSM}).
 
 The results (\ref{gapSM}), (\ref{condSM}) refer to the Schwinger
model with 2 flavors. The Schwinger model with larger number of flavors
also exhibits a critical behavior at $T = m = 0$, and the value of the 
non-universal
critical amplitude for the fermion correlator at any $N$ has been quoted in 
(\ref{A2psiN}). The problem lies, however, in a conformal part of derivation.
The effective low-energy lagrangian for the multiflavor Schwinger model
is
  \be
  \label{SMN}
 {\cal L}^{eff}_N \ = \ \sum_{i=1}^{N} \frac 12 (\partial_\mu
\phi_i)^2 \ - \ C \sum_{i=1}^N \cos \left( \sqrt {4\pi}
\phi_i \right), \nonumber \\
\sum_{i=1}^N \phi_i \ =\ 0
  \ee
  ($C \to 0$ in the
massless limit). To the best of our
knowledge, the model (\ref{SMN}) is not exactly solved, the exact $S$
-- matrix is not known, and the thermodynamic Bethe ansatz technique used
in \cite{Fateev,ZamLesh} to derive the universal relations between
the critical coefficients cannot be applied.

The exact results
(\ref{gapSM}) and (\ref{condSM}) should be confronted with numerical lattice 
simulations. The
path integral calculations in a theory with light fermions are, of course,
much more difficult that in a bosonic theory (like the Ising model), but it
is exactly what is needed in standard 4-dimensional $QCD$. On the other
hand, two-dimensional calculations are much simpler than 4-dimensional ones,
and, if the exact results for the Schwinger model would be reproduced in
such a numerical calculation, the methods to calculate the fermion
determinant etc would be effectively checked and there would be much more 
trust in the lattice results for $QCD_4$ which are of physical interest.

 We are aware of only one paper where relevant numerical computations have been
done \cite{Grady}. The critical exponents coincide with theoretical predictions.
The numerical values for the critical amplitudes 
overshoot the theoretical predictions by $\sim$ 25\%  for the correlation length
and by $\sim$ 35\% for the condensate. It would be very
interesting to repeat these calculations with better accuracy 
on larger lattices, with different
numerical algorithms etc. Two--dimensional models are an excellent playground
where all methods of the lattice gauge theory can be tested.

\section*{Aknowledgements}
I am indebted to I. Affleck, C. Cattringer, R. Guida, M. Grady, J. Hetrick,
 Y. Hosotani, 
B. McCoy, M. Moriconi, J. Verbaarschot,
and Al. Zamolodchikov for illuminating discussions and correspondence.
This work has been done under the partial support of INTAS
Grants CRNS--CT 93--0023, 92--283, and 94--2851.

\end{document}